\documentclass[aps,twocolumn,groupedaddress,showpacs]{revtex4}
\usepackage{graphics}
\usepackage{epstopdf}
\usepackage{soul}
\begin{document}
\newcommand{\beq}{\begin{equation}}
\newcommand{\eeq}{\end{equation}}
\bibliographystyle{apsrev}

\title{Decoherence and Matter Wave Interferometry}
\author{Tabish Qureshi}
\email{tabish@jamia-physics.net}
\affiliation{Department of Physics, Jamia Millia Islamia, New Delhi-110025,
India.}
\author{Anu Venugopalan}
\email{anu.venugopalan@gmail.com}
\affiliation{Centre for Philosophy and Foundations of Science, Darshan Sadan, E-36, Panchsheel Park, New Delhi-110011, India.}
\affiliation{School of Basic and Applied Sciences,
G.G.S. Indraprastha University, Delhi - 110 006, India.}


\begin{abstract}

A two-slit interference of a massive particle in the presence of environment
induced decoherence is theoretically analyzed. The Markovian Master
equation, derived from coupling the particle to a harmonic-oscillator heat
bath, is used to obtain exact solutions which show the existence of  an
interference pattern. Interestingly, decoherence does not affect the pattern,
but only leads to a  reduction in the fringe visibility.


\end{abstract}
\maketitle

\section{Introduction}

Recently it has been possible to observe diffraction of large molecules
like $C_{60}$, which are expected to behave like classical particles
\cite{c60}.  Interference effects have also been observed in larger
molecules \cite{c70}. These effects are a consequence of the linear
superposition principle applied to the wave functions of the particles
and can only be described quantum mechanically. For massive particles,
the existence of such superpositions is classically unimaginable
and is not a familiar observation in the real physical world . The
quantum-classical transition and the nature of classicality as an emergent
property of an underlying quantum system has been the subject of a lot
of study in recent years \cite{zurek,zeh}. The  decoherence phenomenon
has been widely discussed and accepted as the mechanism responsible
for the emergence of familiar classical features in the real  physical
world. Decoherence results from the irreversible coupling of the system to
its environment. The recent experimental observation of  diffraction and
interference patterns for  large molecules raises  a  natural question:
How far  can one go before decoherence effects  destroy the interference
pattern of massive objects? The effect of decoherence on matter wave
interferometry has been explored quite well experimentally. Hornberger
et al have studied the effect of decoherence due to collisions of the
interfering particle with gas molecules \cite{collision}. They  observed
that with increasing pressure of the gas, the fringe visibility goes
down. Chapman et al have studied the effect of photons scattering off
interfering atoms \cite{photon-dec1}. Kokorowski et al have studied the
same effect using multiple photons \cite{photon-dec}. Sonnentag and 
Hasselbach have studied effect of decoherence in electron biprism
interferometer \cite{electron}.
On the theoretical side there have been several focussed studies on
specific systems \cite{electron1, Hornberger}.
However, there is a need to understand the effect of decoherence on
the observed interference pattern in a double-slit experiment in a
simplified way. It is
intuitively obvious that the resulting interference pattern in such a
scenario would  be a consequence of the interplay between the strength
of decoherence, the slit separation and the distance the particle travels
from the slit to the screen. One would like to understand to what extent
each of these parameters play a role.

Decoherence effects on quantum superpositions have generally been
studied  by assuming an initial state which is a superposition of two
spatially localized wave-packets. For an initial state of the form $\psi(x) = \psi_1(x) + \psi_2(x)$, the density matrix in the position representation is

\begin{eqnarray}
\rho(x,x') &=& \psi_1(x)\psi_1^*(x') + \psi_2(x)\psi_2^*(x') \nonumber\\
           &&  + \psi_1(x)\psi_2^*(x') + \psi_2(x)\psi_1^*(x').
\end{eqnarray}
 For localized wavepackets (which best describe a massive particle),
the four terms in this density matrix correspond to four peaks. The last
two terms correspond to the off-diagonal peaks. Decoherence arguments
show that the off-diagonal peaks of the density matrix die out, in time,
because of the effect of the environment \cite{zurek}. Thus, the
appearance of classical behaviour via decoherence is marked by the
dynamical transition of the pure density matrix to a statistical
mixture. For example, using the Markovian Master equations with some
approximations, Zurek has argued that the density matrix for a free
particle in an initial coherent superposition of two Gaussian wave
packets separated by $\Delta x$ decoheres (i.e. the off-diagonal peaks
decay) over a time scale which goes inversely as the square of the
separation, $\Delta x^{2}$, between the two parts of the
superposition \cite{zurek}. For classical systems  and standard
macroscopic separations, this decoherence time is shown to be almost
$10^{-40}$ times smaller than the thermal relaxation time of the system.
Macroscopic superpositions, are, thus, almost instantaneously reduced
to a statistical mixture, a situation which is classically
interpretable. The decoherence approach has been used to study many
models in the context of quantum measurement and the decoherence
mechanism has been explored in the experimental regime also \cite{brune,
cat}. It is now generally accepted that the two main signatures of the
decoherence mechanism are, (i) in the classical regime decoherence takes
place over a time scale that is much smaller than the thermal relaxation
time of the system, and (ii) the decoherence time goes inversely as the
square of the separation between the two parts of the
superposition \cite{av, zurek}. If one were to look at a two-slit
interference situation, one can write a state of the form $\psi(x) =
\psi_1(x) + \psi_2(x)$, where $\psi_1(x)$ and $\psi_2(x)$ correspond to
the probability amplitudes for the particle to pass through slit 1 and
slit 2, respectively. As is well known, the interference pattern
corresponds to the position probability distribution of the time evolved
wave function:
\begin{eqnarray}
|\psi(x,t)|^2 &=& \psi_1(x,t)\psi_1^*(x,t) + \psi_2(x,t)\psi_2^*(x,t) \nonumber\\
           &&  + \psi_1(x.t)\psi_2^*(x,t) + \psi_2(x,t)\psi_1^*(x,t) \nonumber\\
           &=& \rho(x,x,t).
\end{eqnarray}
One might be naively tempted to look at the off-diagonal components of the density
matrix at the screen. However, the interference pattern is obtained from the
probability distribution of of the particle on the screen, which is just the
{\it diagonal} components of the density matrix.
So, it is not obvious if the dying out of {\it off-diagonal} components of the
density matrix, at the screen, as a consequence of decoherence also corresponds
to a disappearance of the interference pattern.
There is a need, therefore, to study the evolution of the state of the
particle along with the effect of the environment, and analyze the
emerging position probability distribution for the existence of an
interference pattern. Savage and Walls have addressed this issue by
studying the evolution of a superposition of two plane-waves under
the influence of decoherence, and its effect on the interference
\cite{savage}. However, though their results illustrate the effect of
decoherence on the interference pattern, their use of plane waves to
describe the state of the interfering molecules seems a little unrealistic
as it is obvious that large  molecules would  be better described by
localized wave-packets.

Recently, decoherence effects on two-slit diffraction has also been
theoretically analyzed by treating the  effect of the environment
using certain phenomenological models \cite{sanz,pascazio}. However,
a fully quantum mechanical analysis, using a microscopic model of the
environment has not yet addressed this issue. This kind of analysis is
very relevant from the point of view of quantitatively probing the elusive
quantum-classical boundary. The study is also highly relevant in the light
of several recent proposals to exploit purely quantum mechanical features
for quantum computation and quantum information processing \cite{ike}.

 In the following, we present an analysis of matter-wave interferometry
 in the presence of a dynamical, quantum environment. Starting with an
 initial superposition describing the double slit situation, we study
 the dynamics of the system through the Markovian master equation which
 takes into account the coupling to the environment. Exact solutions
 for the position probability distribution clearly bring out the role
 of environment-induced decoherence on the interference pattern. The
 rest of the paper is organized as follows. In Sec. II we set up the
 theoretical frame work for our analysis of decoherence and present our
 results. Further, we discuss fringe visibility and some of the factors
 affecting it. Finally, in Sec. III  we summarize the main conclusions
 of this work.

\section{Theoretical Analysis}

\subsection{Coupling to the environment}

The effect of decoherence on the quantum evolution of a system can be
studied by coupling it to a model environment. A popular model for the
environment is a set of non-interacting harmonic oscillators, which may
arise out of different physical situations. The Hamiltonian for a ``free"
particle, coupled to such a model environment can be written as
\begin{equation}
H = {{p}^2\over 2m} + \sum_k \frac{P_k^{2}}{2M_k} + \frac{M_k\Omega_k}{2}\Big(X_k - \frac{C_k x}{M_k\Omega_k^{2}} \Big)^{2} . \label{H}
\end{equation}
Here $x$ and $p$ denote the position and momentum of the particle of
mass $m$, and the second term represents the Hamiltonian for the bath of
oscillators  (environment). $X_k$ and $P_k$ are the position and
momentum coordinates of the $k$th harmonic oscillator of the bath,
$C_k$s are the coupling strengths and $\Omega_k$'s are the frequencies
of the oscillators comprising the bath \cite{legget}. As one is
interested in the dynamics of the particle, and not the detailed
dynamics of the environment (which is not within one's control anyway),
it is conventional to look at the reduced density matrix of the system,
where environment variables have been traced over.

This Hamiltonian has been studied extensively,
to obtain the dynamics of the reduced density matrix. Of particular relevance
here is the case of ``ohmic coupling", which gives the correct limit of
classical dissipation. For our analysis, we deal directly with the
Markovian master equation for the reduced density matrix of the system
in the position representation, where the environmental degrees of
freedom have been traced out\cite{av-dk,dk, legget}:
\begin{eqnarray}
{d\rho_r(x,x')\over dt} &=& {i\hbar\over 2m}\left({\partial^2\rho_r\over\partial
x^2} - {\partial^2\rho_r\over\partial x'^2}\right)\nonumber\\ 
&&-\gamma (x-x')\left({\partial\rho_r\over\partial x}
- {\partial\rho_r\over\partial x'}\right)
- {D\over 4 \hbar^2}(x-x')^2\rho_r .\nonumber\\  \label{master}
\end{eqnarray}
Here $D = 2m\gamma k_BT$  for a thermal bath. Equation (\ref{master}) has been derived
by assuming the harmonic oscillator environment to be in equilibrium, at
a temperature $T$. The parameter $\gamma$ can be assumed to signify
the strength of the coupling of the particle to the environment - it
has its origin in the coupling strengths $C_k$ appearing in the
Hamiltonian (\ref{H}).

At this stage it might be worthwhile to point out that this model of
decoherence is fairly general, and spans a wide range of physical
situations. For example, decoherence due to a photonic heat bath can be
described by a
particle interacting with the modes of the electromagnetic field, modeled by
harmonic
oscillators. In the case of intereference of electrons passing close to a
conducting plate \cite{electron}, the decoherence can be described
by bosonic excitations of
a Fermi sea of conduction electrons - this again, can be done easily by
using a
harmonic oscillator heat bath. In addition to this, the Master equation
(\ref{master}) can also be arrived at by studying the the quantum
evolution of a particle undergoing collisions with smaller
particles, using a stochastic formalism \cite{dk}. This relates very well to
the experimental study of collisional decoherence in matter-wave
interferometry \cite{theory-dc}.

\subsection{Decoherence }

Now that the framework of our analysis is set up, let us get back to
matter wave interferometry. The particle encounters the double-slit,
after which it travels a distance $L$, say, along the y-direction,
before registering on the screen. Clearly, for the  interference pattern
to be visible, coherence along the x-direction is important, whereas the
dynamics along the y-axis just serves to transport the particle from the
slits to to the screen. In order to simplify the calculations, we assume
that the particle emerges from the double-slit, travels along the y-axis
with a well-defined average momentum $p_0$ and reaches the screen after
a time $t_L = mL/p_0$.  We now focus only on the time evolution of the
particle's wave function in the x-direction.

At this stage, let us  specify the functional form of the initial state
that emerges from the double-slit.
We assume that the action of the slit is to prepare a superposition
of two Gaussian wave-packets, centered at the location of the respective slits,
with a width equal to the width of the slit. We define the initial state as
\begin{eqnarray}
\psi(x,0) &=&  {1\over\sqrt{2}}{1\over (\pi/2)^{1/4}\sqrt{\epsilon}}\left(
e^{-(x-x_0)^2/\epsilon^2} + e^{-(x+x_0)^2/\epsilon^2}\right),\nonumber\\
\label{slitstates}
\end{eqnarray}
where $x_0 = d/2$, $d$ being the separation between the two slits. (5)
represents two Gaussians centered at $x = \pm x_0$. 

It may be noted that we have ignored the decoherence effects that could occur
{\em before} the particle reaches the slit. From the point of
view of an analytical calculation, it is  difficult to take into account
the effect of decoherence both before and after the double slit. Moreover,
there could be specific experimental situations where decoherence occurs
only after the double slit. In the experiment of Sonnentag and Hasselbach
on interference of electrons, the decoherence takes place due to a
metallic plate kept after the double-slit \cite{electron}.

For the time evolution after the double slit, the initial density matrix for the particle state can be written using (\ref{slitstates}), as 
\begin{eqnarray}
\rho(R,r,0) &=& {1\over \sqrt{2\pi\epsilon^2}}\left( e^{-{(R-d)^2+r^2\over 2\epsilon^2}} +  e^{-{(R+d)^2+r^2 \over 2\epsilon^2}}\right. \nonumber \\
&&\left. +  e^{-{(r-d)^2+R^2\over 2\epsilon^2}}
+ e^{-{(r+d)^2+R^2\over 2\epsilon^2}}\right),  \label{rho0}
\end{eqnarray}
where $R=x+x'$ and $r=x-x'$ and $d$ is the slit separation. 
In order to analyze interference, one needs to obtain the density matrix of the particle at a time $t_L$, when it reaches the screen. For this, we solve the master equation (\ref{master}), with the initial condition (\ref{rho0}).
It turns out that an exact solution for $\rho_r(x,x,t_L)$ can be obtained.
This will contain full information about the interference and the degree of decoherence. However, for this purpose we do not need the full density matrix of the particle - we are
only interested in the position probability distribution of the
particle on the screen, which is given by the diagonal part ($x=x'$) of the
reduced density matrix. From the exact solutions one can see that the
position probability distribution of the particle on the screen has the
final form: 
\begin{eqnarray}
 \rho_r(x,x,t_L) &=& {1\over\sqrt{\pi}\Omega}\left({1\over 2} e^{-(x-x_0)^2\over\Omega^2}
+ {1\over 2} e^{-(x+x_0)^2\over\Omega^2} + \right. \nonumber\\
&& \left.  e^{-{x^
2+x_0^2\over\Omega^2}}e^{-{\Gamma x_0^2
\over\Omega^2}}
\cos\left\{{xd\hbar(1-e^{-2\gamma t_L})\over m\gamma\epsilon^2\Omega^2}
\right\}\right),
\label{psisq} \nonumber\\
\end{eqnarray}
where $\Gamma = {D\over 16m^2\epsilon^2\gamma^3}(4\gamma t_L+4e^{-
2\gamma t_L}
-e^{-4\gamma t_L} - 3)$, $\Omega^2=\epsilon^2+{\hbar^2(1-e^{-2\gamma t_L})^2\over
m^2\epsilon^2\gamma^2} + \Gamma $.

Starting from (\ref{master}), the result (\ref{psisq}) is
exact and represents the position probability distribution at the screen. The presence of the cosine term indicates the existence of the interference pattern. However, (7) is a very general result which can be used to study many things, like damped motion of the particle, the effect of friction and of course, decoherence. In order to relate it to decoherence, this result
should be analyzed in a suitable limit. This limit is set by the requirement
that the interaction of the particle with the environment be so weak that
the dissipative effects are not noticeable, only the decoherence is.
This will correspond to the limit $1/\gamma \gg t_L$, i.e, when the
relaxation time of the system is much much larger than the 
time scales over which the experiment is performed and the observations
made. In the following we make this approximation:
\begin{eqnarray}
\Omega^2 &\approx& \epsilon^2 + \left({2\hbar t\over m\epsilon}\right)^2
 + {2Dt_L^3\over 3m^2} \approx {\lambda_d^2L^2\over\pi^2\epsilon^2}, \nonumber\\
\Gamma &\approx& {
2Dt_L^3\over 3m^2 \epsilon^{2}} = {D\lambda_d^2L^2\over 6\pi^2\hbar^2}t_L.
\label{limit}
\end{eqnarray}
Here, we have assumed that $t_L = mL/p_0$, which implies that
$\hbar t_L/m = \lambda_d L/2\pi$, $\lambda_d$ being the de Broglie wavelength. 
In the expression for $\Omega^2$, we have made a further approximation,
$\epsilon \ll \lambda_dL/\pi\epsilon$, i.e., the spread of the wave-packets,
after travelling a distance $L$, is much larger than the original width.
This will happen when the original width is very small. This is a realistic expectation since the slits should be narrow enough to let  the two wave-packets
 spread and overlap with each other to lead to an  interference pattern on the screen.

Using the approximations (\ref{limit}), (\ref{psisq})
reduces to
\begin{eqnarray}
\rho_r(x,x,t_L)
&=& {1\over\sqrt{\pi}\sigma}\left({1\over 2} e^{-(x-x_0)^2\over\sigma^2}
+ {1\over 2} e^{-(x+x_0)^2\over\sigma^2} + \right. \nonumber\\
&& \left.  e^{-{x^2+x_0^2\over\sigma^2}}\exp(-{t_L \over 24\tau_D})
\cos\left\{{\pi xd\over \lambda_d L}.
\right\}\right),\nonumber\\
\label{pattern}
\end{eqnarray}
where  $\sigma = \lambda_dL/\pi\epsilon$ and 
$\tau_D =  {\hbar^2\over 2m\gamma k_BTd^2}$. Recalling earlier results
on decoherence, one can recognize $\tau_D$ as the decoherence time of a superposition of two wave-packets, separated by a distance $d$, due to interaction with an environment at temperature $T$, with a coupling strength, or relaxation
rate $\gamma$\cite{zurek, av}.

Note that without the term $\exp(-{t_L \over 24\tau_D})$, (\ref{pattern}) represents the interference (position probability distribution) of two wave-packets of initial width $\epsilon$ each. The expression also represents the interference pattern corresponding to a matter-wave of de-Broglie wavelength $\lambda_d$, having travelled a distance $L$ from the double slits. The decoherence effects come in only through the term $\exp(-{t_L \over 24\tau_D})$. This term affords a 
simple physical meaning - two wave-packets, separated by the slit distance
$d$ lose their coherence after a time $\tau_D$, which, by definition, is the
decoherence time. If the wave-packets reach the screen at a time $t_L$ which
is much much larger than $\tau_D$, no interference will be visible. For
the interference pattern to be observable, $\tau_D$ should obviously be of the
order of  $t_L$.

Since our analysis is not tied to any one experimental situation, we use
the following arbitrary values for various parameters:
$d = 1~\mu$m,
$\epsilon = 0.1~\mu$m, $\lambda_d = 5\times 10^{-6} \mu$m, L = 20 cm.
We assume that the Gaussian width of a wave-packet emerging out of a
rectangular slit of width $w$ will be roughly $w/2$. The dimensionless
parameter $t_L/\tau_D$ is a measure of decoherence.
Using these values, expression
(\ref{pattern}) is plotted in FIG. 1. Note that the overall Gaussian
profile is due to the finite width of the slits. If the slits were infinitely
narrow, one would see a flat profile with narrow interference peaks.

\begin{figure}
\resizebox{6.5cm}{!}{\includegraphics{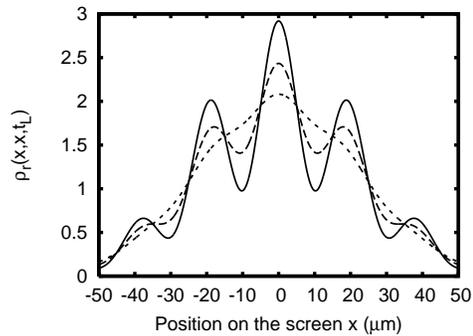}}
\caption{The interference pattern $\rho_r(x,x,t_L)$ plotted against the position
on the screen $x$, for $\epsilon = 0.1 \mu m, \lambda_d = 5\times
10^{-6} \mu$m, $L = 20$ cm,  $d=1.0\mu$m, and for various values of
$t_L/\tau_D$. Solid line represents
$t_L/\tau_D=4.0$, dashed line represents $t_L/\tau_D=20$ and
the dotted line represents $t_L/\tau_D=60$.}
\end{figure}

\subsection{Fringe visibility}

One quantity of particular importance in matter-wave interferometry is the fringe visibility, and the effect of decoherence on it. Conventionally, it is defined as 
\begin{equation}
{\cal V} = {I_{max} - I_{min} \over I_{max} + I_{min} } ,
\end{equation}
where $I_{max}$ and $I_{min}$ represent the maximum and minimum intensity
in neighbouring fringes, respectively. In reality, fringe visibility will
depend on many things, including the width of the slits. For example, if
the width of the slits is very large, the fringes may not be visible at all.
If we focus on just the effect of decoherence on fringe visibility, we
can assume that the slits are so narrow that we get an essentially flat
background profile. This means that $\sigma$ will be so large that the
functions $e^{-(x \pm x_0)^2\over\sigma^2}$ will have the same values at the
points of maximum and minimum intensity. Maxima and minima of (\ref{pattern}) will occur at points where the argument of the cosine is 1 and -1, respectively.
Taking two such neighbouring points, the  fringe visibility can be written as
\begin{equation}
{\cal V} = {\exp(-{t_L \over 24\tau_D})
 \over \cosh({2x_n x_0\over\sigma^2}) } , \label{visibility}
\end{equation}
where $x_n$ denotes the mean position of $n$th fringe. Clearly, the fringe
visibility goes down as $t_L$ becomes larger than $\tau_D = 
{\hbar^2\over 2m\gamma k_BTd^2}$. This can most easily happen when either
$\gamma$ or $m$ becomes large.

The expression for fringe visibility, (\ref{visibility}), details its
dependence on various physical parameters. It can be written in an expanded
form as follows:
\begin{equation}
{\cal V} = {\exp(-{m\gamma k_BTd^2t_L \over 12\hbar^2})
 \over \cosh({2x_n x_0\over\sigma^2}) } . \label{visibility1}
\end{equation}

If the present microscopic model is compared with certain stochastic models
of the environment, the parameter $\gamma$ turns out to be proportional to the
rate of collision of the interfering particle with the smaller particles
constituting the environment \cite{dk}. If one tries to relate this model
to collisional decoherence, as in the experiment of Hornberger et. al.
\cite{collision}, the pressure of
the gas in the chamber should be directly proportional to the collision
rate, and therefore to the parameter $\gamma$ in our calculation. Eqn.
(\ref{visibility1}) indicates an exponential decay of visibility with
$\gamma$. This implies that the fringe visibility should go down
exponentially with increasing pressure. This is in broad agreement with the
experimental findings of Hornberger et. al. \cite{collision}, and does not
depend on the fact that Hornberger et. al. use a Talbot-Lau interferometer
instead of a double-slit.
If one relates the calculation here to the experiment of Sonnentag and
Hasselbach on interference of electrons, the decoherence is due to the
electron's interaction with the metallic electrons of the plate. In this
case, $\gamma$ is given by \cite{electron}
\begin{equation}
\gamma = {e^2 \rho \over 32\pi m z^3} ,
\end{equation}
where $\rho$ is the resistivity of the place and $z$ is the distance of the
interfering beam from it. So, in this case too, the decoherence effects are
expected to be stronger when electrons are closer to the plate, or $z$ is
smaller. This would mean, larger value of $\gamma$.

Fringe visibility is also very sensitive to the separation between the
two slits, $d$. It decays exponentially with the square of the slit
separation. To get an idea of the magnitude of the effect, for example,
if a particular slit separation gives a visibility of 60 percent,
doubling the slit separation will reduce the visibility to about 13
percent. 

It is also clear from (\ref{visibility1}) that fringe visibility goes down
exponentially with the mass of the interfering particle.
So, if one were to use a molecule, say, twice as heavy as $C_{60}$, one would
have to either reduce the temperature by the same factor, or decrease the
pressure by the same factor, in order to get the same visibility as for
$C_{60}$. This is in tune with the general expectation that the interference
arising from the quantum nature of particles in the double-slit scenario
analyzed here will be more vulnerable to decoherence for more massive
particles. In many situations, the coupling to the environment may
also be related to the mass, and such simplified logic may not always be
correct.

\section{Conclusion}
We have theoretically analyzed the effect of decoherence on a massive
particle, in a two-slit interference set-up, using a quantum dynamical model of the environment. Interestingly, the interference pattern does not get distorted
as a result of decoherence. The effect of decoherence is only to reduce the
visibility of the interference fringes. The fringe visibility crucially
depends on the decoherence time of a superposition of two freely evolving
wave-packets, initially localized at the centers of the two slits. Our
results clearly demonstrate the two main signatures of the decoherence
mechanism, namely (a) the decoherence time is much smaller than the
thermal relaxation time, and (b) the decoherence time is inversely
proportional to the square of the ``separation'' between the two parts
of the superposition. Our results also  show that the fringe visibility
goes down exponentially with the mass of the particle, with the
temperature of the environment, with the square of the slit separation
and with the coupling strength of the
particle to the environment. Finally, the results are in good qualitative
agreement
with the matter-wave interferometry experiments carried out to study
decoherence effects with large molecules and also with electrons.

\end{document}